\documentclass[aps,prd,twocolumn,superscriptaddress,nofootinbib,
nobalancelastpage,showpacs,nobibnotes]{revtex4}

\usepackage{epsfig}

\begin{document}
\newcommand{\identity}{\:\mbox{\sf 1} \hspace{-0.37em} \mbox{\sf 1}\,}

\title{State permutations
from manipulation of near level-crossings.}
\author{Nicole F. Bell}\email{nfb@fnal.gov}
\affiliation{NASA/Fermilab Astrophysics Center, Fermi National
Accelerator Laboratory, Batavia, Illinois 60510-0500}
\author{R. F. Sawyer}\email{sawyer@vulcan.physics.ucsb.edu}
\affiliation{Department of Physics, University of California at
Santa Barbara, Santa Barbara, California 93106}
\author{Raymond R. Volkas}\email{r.volkas@physics.unimelb.edu.au}
\affiliation{School of Physics, Research Centre for High Energy
Physics, The University of Melbourne, Victoria 3010, Australia}
\author{Yvonne Y. Y. Wong}\email{ywong@physics.udel.edu}
\affiliation{Department of Physics and Astronomy, University of
Delaware, Newark, Delaware 19716}

\begin{abstract}
We discuss some systematic methods for implementing state
manipulations in systems formally similar to chains of a few spins
with nearest-neighbor interactions, arranged such that there are
strong and weak scales of coupling links. States are permuted by
means of bias potentials applied to a few selected sites. This generic 
structure is then related to an atoms-in-a-cavity model that has been
proposed in the literature as a way of achieving a decoherence free
subspace. A new method using adiabatically varying laser
detuning to implement a CNOT gate in this model is proposed. 
\end{abstract}

\pacs{3.65Xp, 3.67a \hfill FERMILAB-Pub-02/057-A}

\maketitle

\section{Introduction}

Two central concerns of quantum information theory are
representation of data through occupancy of quantum states and the 
manipulation of these data. Roughly speaking, for these 
purposes, we desire systems that have the two properties:

1) When the system is left alone it should have some 
eigenstates that are quite localized, in the sense that 
the occupancy of such a state identifies a property of 
a localized physical element (e.g., a particular spin 
being up or down, or a state of a single atom being in 
one energy level or the other. )

2) There is a simple mechanics for applying a signal 
from the outside that will move the system from one 
of these ``localized'' states to another.

In this paper we discuss two models that show some 
interesting potentialities for this program. The first is a novel 
and rather idealized ``spin chain'' type of model; the second 
is an ``atoms-in-a cavity" model, still quite idealized, that has 
been discussed at some length in several previous works by 
other authors. 
 
All of the models that we shall consider (and some others in 
the literature) have a unifying feature that can be stated as a 
nearly-general property of a class of Hamiltonians. We define 
this class as follows:

a) We picture the states as points distributed in a space, and 
connect each pair of points that correspond to a 
non-vanishing matrix element of $H$. We consider only cases
in which each point is connected to a few neighbors.
A natural extension that will be required for the
atoms-in-a-cavity discussion is to include the
``environment'' as one of the points in the space, with
dissipative links to it from some of the other states.

b) Now let there be two scales of strength in these 
connections, strong (S) and weak (W). If there are small 
neighborhoods, or blocs, in this space in which all 
connections, as well as the diagonal elements, are weak, then 
there will generally be energy eigenfunctions that are nearly 
confined to these blocs. This can be seen by treating the weak 
links to the outside perturbatively after diagonalizing the 
weak bloc in the absence of connections to the outside. Since 
the denominators in this perturbation are generally of strong 
order and the numerators of weak order, the perturbation will 
make little difference to the state. We can characterize
the perturbation of the state as being of order ${\rm W/S}$.
Likewise, we note that 
although other weak blocs somewhere else in the space of 
states will indeed have energies of weak scale, we are in 
general protected from a small denominator by the fact that 
we must go through a weak link to get from the first bloc into 
an intervening strong bloc, and then another weak link to get 
to the second weak bloc, where we would find energies of the 
weak order characterizing the first bloc. Thus the largest term
with a dangerous denominator would generally be of order 
${\rm W^2/SW}$. (The dissipative connections to the environment in the
atoms-in-a-cavity model to be considered later will come
to be classified as, effectively, a set of additional strong links.)

The above statement is, to be sure, neither a theorem nor an 
unfamiliar kind of conclusion. In much more sophisticated 
work on infinite systems much is known about the relations 
between disorder and localization. But we will show that the 
result above, tempered as it is with ``generally'' (meaning
``most of the time'') is a very useful design criterion for 
systems that meet the objectives enunciated at the beginning
of this paper. 

We will begin by considering cases where the strong bloc consists of 
two states connected by a single strong link, such that the Hamiltonian 
is of the form
\begin{equation}
\left(\begin{array}{cc}
0 & S    \\ 
S & 0 
\end{array}\right),
\end{equation}
and the eigenvalues of the two energy eigensates are $\pm S$.  
Adding weak scale connections to other states, will make only a small 
perturbation to these eigenstates.
We may generalise this to chains of an arbitrary {\it even} number 
of states connected by strong links of similar size -- all the eigenvalues 
will be of order strong.  However, if we have an {\it odd} number of 
states, there will always be a zero eigenvalue, for example, three states 
connected via a Hamiltonian of the form
\begin{equation}
\left(\begin{array}{ccc}
0 & S_{12} & 0   \\ 
S_{12} & 0 & S_{23} \\
0 & S_{23} & 0 
\end{array}\right),
\end{equation}
will have eigenvalues $\{ 0, \pm \sqrt{S_{12}^2 + S_{23}^2}  \}$.  Due to the 
presense of a zero eigenvalue, adding weak scale links to states outside 
of this bloc will greatly perturb the eigenstates. 
We regain of a set of eigenvalues, all of strong order, however, if we
introduce either a coupling $S_{13}$ between the 1st and the 3rd 
states, or diagonal elements of strong scale.  Dissipative connections, for 
example, would enter as imaginary diagonal terms.

Once we have reason to take a weak bloc as isolated, we then 
wish to address the movements of probability through the 
action of time dependent external interventions. In the first  
chain-of-spins model that we shall present, with five sites, 
this will take the form of a bias signal to be applied to the 
middle site, which is changed slowly enough for adiabaticity 
of a kind to be realized. The actual state transformations that 
we discuss can be regarded as fairly 
elaborate ``avoided-crossing'' phenomena. 
In the atoms-in-a-cavity models of 
Refs.\ \cite{beige1}, \cite{beige2}, and 
\cite{pachos} that we consider, after noting how the
creation therein of a decoherence-free subspace
can be qualitatively understood using the strong-weak
picture, we 
demonstrate an adiabatic method of manipulation by tuning 
and detuning that is different from the pulsed, and 
exactly-timed, laser techniques that are suggested 
in these references.

The remainder of this paper is structured thus: The
next section analyses spin-chain models with a
W and S hierarchy of nearest-neighbor couplings
and time-dependent biases on selected sites.
Section \ref{atomsinacavity} then re-examines 
atoms-in-a-cavity models, while Sec.\ \ref{conclusion}
provides further discussion and a conclusion.

\section{Spin-chain models}
\label{spinchain}

Our first example uses a chain of five spins. We 
take the spins to be numbered
consecutively from left to right, with pure exchange interactions
built of the operators between adjacent spins $i$ and $j$,
\begin{equation}
h^{(i,j)}=\sigma_+^{(i)}\,\sigma_-^{(j)}+\sigma_-^{(i)}\,\sigma_+^{(j)}.
\end{equation}
We choose the Hamiltonian
\begin{equation}
H=g_1 h^{(1 , 2)} +\lambda h^{(2 , 3)}+\lambda h^{( 3,4 )}+g_2 h^{( 4, 5)}
+ f(t)[\sigma_z^{(3)} - 1],
\label{ham}
\end{equation}
where we have added a single time-dependent term involving the
operator of the middle spin $\vec \sigma^{(3)}$. We refer to the
function $f(t)$ as the bias on site \#3. Since this Hamiltonian
commutes with the operator $\sum_i \sigma_z^{(i)}$, we can operate
within the set of five states with four of the spins up and one
spin down. The S-W structure comes from 
taking $\lambda \ll g_1,g_2$, and $g_1 \neq g_2$. When
$f(t)=0$, this choice has the effect of creating eigenstates of
$H$ that are very nearly the following,
\begin{eqnarray}
\psi_1&=&(\uparrow \downarrow +\downarrow \uparrow  )(\uparrow
\uparrow \uparrow )/\sqrt{2}, \nonumber\\ \psi_2&=&(\uparrow
\downarrow -\downarrow \uparrow  )(\uparrow \uparrow \uparrow
)/\sqrt{2}, \nonumber\\ \psi_3&=  &(\uparrow \uparrow \downarrow
\uparrow\uparrow  ), \nonumber\\ \psi_4&=&(\uparrow \uparrow
\uparrow) (\uparrow \downarrow +\downarrow \uparrow  )/\sqrt{2},
\nonumber\\ \psi_5&=&(\uparrow \uparrow \uparrow )(\uparrow
\downarrow -\downarrow \uparrow  )/\sqrt{2}.
\label{5states}
\end{eqnarray}
We understand these combinations qualitatively by noting that they
would be the exact $f=0$ eigenstates if the small coupling
$\lambda$ were set to zero, and that, treating $\lambda$ as a
perturbation, the link between the pairs of states (\#1,\#2) and
(\#4,\#5) is of order $\lambda^2$.

Now we ask the question: Of the 120 permutations on this set of
five states, how many can we implement by applying a small
series of simple pulses in $f(t)$? In forming these pulses we
shall tailor $f(t)$ to the permutation that is sought, but we
shall insist that $f(t)$ begins at and ends at the value zero. By
a permutation we mean just the reshuffling of the
states, modulo phase. This demand effectively rules out setting
relative phases, which would require fine-tuning in any system
that is to be considered over a period of time.
The answer to the question is ``all 120''.   

The method uses adiabatic avoided-level-crossing
dynamics, with an additional feature that can be embodied in the
above Hamiltonian, namely, that the bias on the weakly coupled
site can be changed suddenly, without affecting the state of the
system appreciably, at all times when the system is far from any
(near) level-crossing. As an example, we choose the parameters
$g_1=30$, $g_2=60$, $\lambda=1$, and begin by defining two bias
operations $f(t)$,
\begin{eqnarray}
&f_a(t) = (t-t_0) \,\, \theta[t-t_0]\theta[t_0+\tau-t]
\rightarrow U_a(t_0), \nonumber\\ &f_b(t)=-f_a(t) \rightarrow
U_b(t_0). \label{biases1}
\end{eqnarray}
The signal begins at $t=t_0$, grows linearly until $t=t_0+\tau$ 
when it is switched off abruptly. We take $\tau=20$ in
the applications that follow. Any bias $f_{\alpha}(t)$
defines a transformation,
\begin{equation}
U_{\alpha}(t_0)=T[\exp(-i\int_{t_0}^{t_0+\tau} dt' H(t'))].
\label{trans}
\end{equation}

We will show \cite{footnote1} that the
operation $U_a$ effects the permutation $\langle 13 \rangle$, 
while $U_b$ effects $\langle 23 \rangle$.
The turn-off of the bias at $t=t_0+\tau$ leaves the state
at time $t=t_0+\tau$ evolving with the Hamiltonian of Eq.\ (\ref{ham})
with $f=0$, so that we are prepared to apply another signal to get
a compound of the permutations. Thus we can write,
\begin{equation}
U_{1 \Leftrightarrow 2}\equiv [U_a(2\tau)][U_b(\tau)][U_a(0)],
\label{compound}
\end{equation}
where the time arguments emphasize that the transformations are to
be executed serially over a total time of $3\tau$. Using the
composition property of permutations $\langle 13 \rangle \langle
23 \rangle  \langle 13 \rangle = \langle 12 \rangle $, we see that
$U_{1 \Leftrightarrow 2}$ just interchanges states \#1 and \#2,
leaving \#3 alone. The basis for anticipating the actions of both
the individual and compound transformations can be shown in an
averted level-crossing diagram based on Eq.\ (\ref{ham}) with the bias
function given by,
\begin{equation}
f(t)=f_a(t)+f_b(t-\tau)+f_a(t-2 \tau).
\label{examplebias}
\end{equation}
Figure 1 shows plots of the energy levels for the first
three states as they evolve under the above successive changes of
the bias.

%%%%%%%%%%%%%%%%%%%%%%%%%%%%%%%%%%%%%%%%%%%%%
\begin{figure}[t]
\begin{center}
\epsfig{file=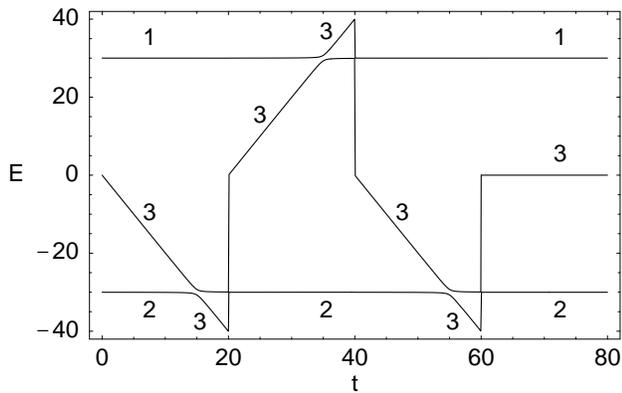,width=3.25in}
\caption{\label{fig1}
Level diagrams for the three states that are appreciably affected
by the pulse chain of Eq.\ (\ref{examplebias}). The numbers on the curves
indicate possible initial states, enumerated as in Eq.\ (\ref{5states}). In the
adiabatic limit, the system, beginning in one of these states,
follows the continuous path connected thereto. For example, if the system
is initially prepared in state \#1, then it first adiabatically transforms
to state \#3, remains there during the sudden bias jump at $t=40$, then
adiabatically transforms to state \#2.
States \#4 and \#5
remain virtually unmixed and at their original energies.
The energy is is units of $\lambda$ and the time in units of $\lambda^{-1}$. }
\end{center}
\end{figure}  
%%%%%%%%%%%%%%%%%%%%%%%%%%%%%%%%%%%%%%%%%%%%%

To show that the states really follow these paths, we construct
the matrix $U(t)=T[\exp(-i\int_{0}^{t} dt' H(t'))]$ by directly
solving the Schr\"{o}dinger equation with the time-dependent bias
of Eq.\ (\ref{examplebias}). In Figs.\ 2 and 3, we plot $|\langle 1,2,3
|U(t)|1\rangle|^2$ and $|\langle 1,2,3 |U(t)|2\rangle|^2$ against
time in this solution, giving the expected behavior, $|U( 3
\tau)|^2=|U_{1 \Leftrightarrow 2}|^2$ with negligible
contamination either from non-adiabaticity in the region of small
level separation, or from transitions induced in any of the three
sudden changes. The plots show clearly the effects of each of the
constituent transitions in turn, giving rise to the factorization
indicated in Eq.\ (\ref{compound}).

\begin {figure}[ht]
    \begin{center}
        \epsfxsize 3.25in
        \begin{tabular}{rc}
            \vbox{\hbox{
$\displaystyle{ \, { } }$
               \hskip -0.1in \null} %\vskip 0.2in
} &
            \epsfbox{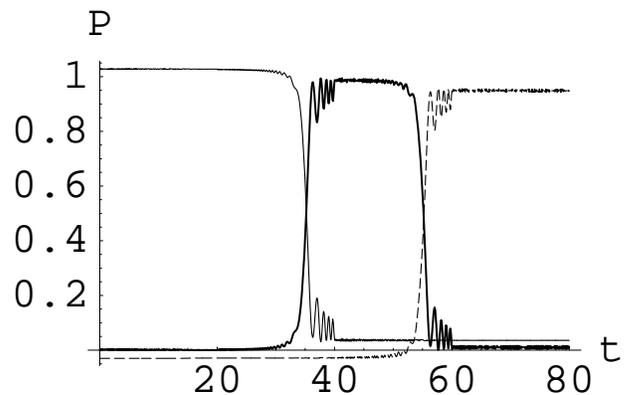} \\
            &
            \hbox{} \\
        \end{tabular}
    \end{center}
\label{fig2}
%\vskip 1in
\protect\caption
    {%
The evolution of probabilities for the Hamiltonian of Eq.\ (\ref{ham})
with $f(t)$ from Eq.\ (\ref{examplebias}), and the initial state \#1.
The light solid curve is the probability of the system in state
\#1 (displaced slightly upward for clarity), the heavy solid curve
is for \#3, and the dashed curve is for \#2 (displaced slightly
downward). The time is given in units of $\lambda^{-1}$.}
\end {figure}

\begin {figure}[ht]
    \begin{center}
        \epsfxsize 3.25in
        \begin{tabular}{rc}
            \vbox{\hbox{
$\displaystyle{ \, { } }$
               \hskip -0.1in \null} %\vskip 0.2in
} &
            \epsfbox{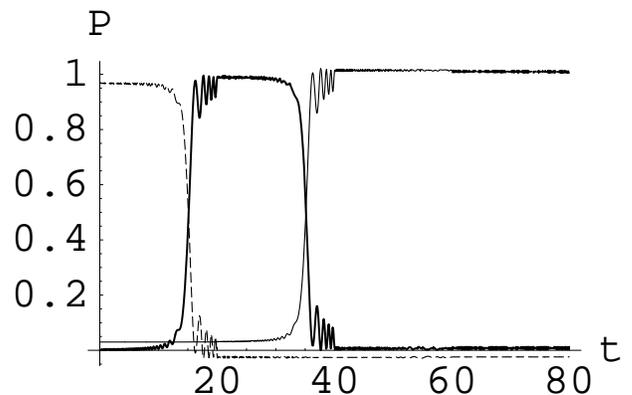} \\
            &
            \hbox{} \\
        \end{tabular}
    \end{center}
\label{fig3}
%\vskip 1in
\protect\caption
    {%
Same as Fig.\ 2, but with the initial state taken as
 \#2. Together, Figs.\ 2 and 3 show the interchange operation
 effected by the particular pulses of (\ref{examplebias}).
The time is given in units of $\lambda^{-1}$.
}
\end {figure}

To implement the transformations involving states \#4 and \#5, we
introduce four more bias functions,
\begin{eqnarray}
f_c(t)&=&2(\tau-t)\theta(\tau-t)\theta(t)\rightarrow
U_c\rightarrow \langle 352 \rangle , \nonumber\\
f_d(t)&=&-f_c(t)\rightarrow U_d\rightarrow \langle 341\rangle ,
\nonumber\\
f_e(t)&=&-2(\tau-t/2)\theta(\tau-t)\theta(t)\rightarrow
U_e\rightarrow \langle 34 \rangle , \nonumber\\
f_f(t)&=&-f_e(t)\rightarrow U_f\rightarrow \langle 35 \rangle ,
\label{biases2}
\end{eqnarray}
where, for simplicity, we have chosen $t_0=0$. By straightforward
multiplication of the operations in Eqs.\ (\ref{biases1}) and
(\ref{biases2}), we obtain $U_dU_a\!\! \rightarrow \! \!\langle14
\rangle$, $U_cU_b \!\! \rightarrow\!\! \langle 25 \rangle$,
$U_aU_fU_a\!\! \rightarrow\!\! \langle15 \rangle$, $U_bU_eU_b\!\!
\rightarrow\! \langle 24 \rangle $, $ U_eU_f U_e \! \! \rightarrow
\!\! \langle 45 \rangle $. These five composite operations,
together with $U_aU_bU_a \! \! \rightarrow\!\! \langle 12 \rangle
$ and the four single pulse operations $U_a,U_b,U_e,U_f$, give all
of the simple interchanges. All permutations can be built from
these interchanges, although in most cases it would be more
efficient to draw on the cycle-of-three permutation operators
$U_c$ and $U_d$ as well, or on further primary pulse shape
variants that  directly embody other operators.

Now we consider longer chains with nearest neighbor
couplings,
\begin{equation}
H_0= \sum_{i=j+1}g_{i,j}h^{(i,j)}-E_0,
\end{equation}
where $g_{i,j}$ are coupling coefficients, and $E_0$ is the energy
that gives $\langle H_0 \rangle =0$ for the case of all spins up.
We again adopt two scales of coupling strength: S and W.
As an example, we consider the pattern of couplings
$\{g_{1,2},\,\,g_{2,3},\,\,g_{3,4}, \ldots \}$ to be \{S, W, W, S,
W, W, S, W, W, \ldots \}. Under this scheme, the states
\#3,\#6,\#9... are only weakly coupled to their neighbors.

As before, we take only states with one spin down and the
remainder up. With the W coupling constants turned off, the
eigenvalues of the states in which the down spin occupies one of
the large blocs, i.e., \{(\#1,\#2), (\#4,\#5), (\#7,\#8) ...\},
come in pairs of $\pm g_{i,i+1}$. We assume that the S couplings
are sufficiently irregular so that the differences in energies
between nearest strong blocs, e.g., $g_{1,2}\!-\! g_{4,5}$, are
still of the strong (S) scale. When we add the W couplings, the
eigenvectors will remain almost localized, as can be seen from a
perturbation expansion. In general, as argued previously,
small energy differences
between two non-adjacent blocs are not a concern, since as we
move across $n$ additional weak links we can tolerate energy
denominators that are smaller by a factor of $(g_W/g_S)^{2n}\ll 1$.

 Thus for most cases the eigenvectors can
be arranged in a list
$\{(\xi_1,\uparrow,\uparrow,\uparrow,...),\,\
(\eta_1,\uparrow,\uparrow,\uparrow,...),\,\
(\uparrow,\uparrow,\downarrow,\uparrow,\uparrow,...),\,\
(\uparrow,\uparrow,\uparrow,\xi_2,\uparrow,...),\,\
(\uparrow,\uparrow,\uparrow,\eta_2,\uparrow,...),\,\
(\uparrow,\uparrow,\uparrow,\uparrow,\uparrow,\downarrow,\uparrow,...),\,\
...\}$. Here, the $\xi$'s and $\eta$'s stand for the symmetric and
anti-symmetric two-component eigenstates of the strong blocs with
one spin up and the other down. If the system begins with all of
its amplitude in one bloc, the amplitude will stay almost within
that bloc under the evolution governed by $H_0$.

Now, generalizing the earlier five-site example, we can ask
whether biases placed at every third site, \#3, \#6, \#9, ...,
which are weakly coupled both to the right and to the left, have
the capability of moving information around the whole system. We
have looked at an eight-site example, with independently
manipulable biases at sites \#3 and \#6. In complete analogy
with the five-site case, we find it is easy, for example, to move
one of the (\#1,\#2) eigenstates successively to the blocs (\#3),
(\#4,\#5), (\#6), (\#7,\#8), again by numerically solving the
Schr\"{o}dinger equation for  the system.

As a qualitative summary of these outcomes, we restate and
sharpen some of the remarks in the introduction:

(1) If we have a chain of blocs of states connected, one to
another, through an intermediary state that is weakly coupled to
both blocs, the dynamics within the blocs will be nearly
self-contained. That is, if the probability is localized within a
particular bloc at a particular time, in the form of any superposition
of the eigenstates belonging to that bloc, then the probability
will remain in that bloc. Likewise, a bloc of several states in
which the mutual couplings and the diagonal elements are weak
will, in general, not admix appreciably with strongly coupled states
to its left or right.

(2) By putting controllable biases on a weak connection site to
bring the energy of an associated level to (nearly) coincide with
a level in an adjacent strong bloc, shifts of probability from
bloc to bloc can be implemented in an orderly and complete
fashion.  

\section{Atoms-in-a-cavity models}
\label{atomsinacavity}

An exact analogue to the above mechanics for moving a 
system from one state to another can be applied to some models
of atoms-in-a-cavity type that have been the subject
of a series of recent papers. As an example, we consider
the model of Ref.\cite{beige1}, with two identical atoms in a cavity,
and a dynamics that is effectively confined to three
states, which form a decoherence free subspace (DFS).

\begin{figure}[t]
\begin{center}
\epsfig{file=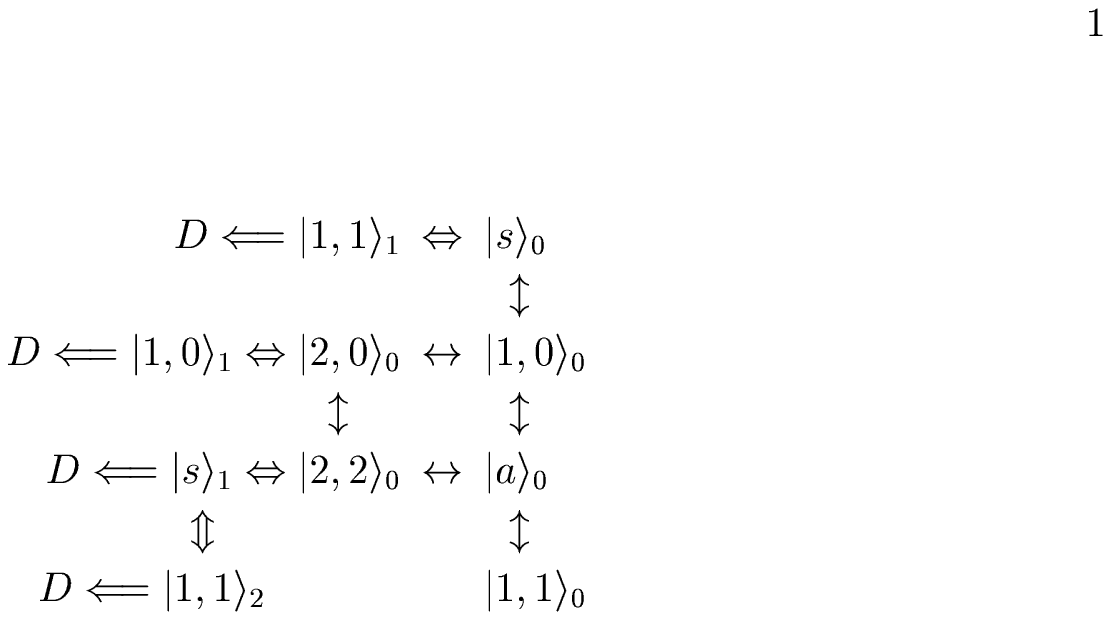,width=3.25in}
\caption{\label{connections}
In a notation close to that of Ref. 
\cite{beige1}, 
the states are
labeled as $|{\rm atom}\,\# 1,{\rm atom}\,\# 2\rangle_{\rm cav}$
where ``cav'' is the excitation of the cavity mode. The states
denoted ``s'' and ``a'' (for the atomic part) are $| {\rm s,a}
\rangle =(|1,2 \rangle \pm |2,1 \rangle )/\sqrt{2}$. The
connections are those variously induced by the three laser
couplings and the cavity mode, as explained in the text.}
\end{center}
\end{figure}

In this model, the two atoms have three levels, $\{0,1,2\}$, and are 
placed in a cavity tuned exactly to the $1\!-\!2$ level spacing, with the 
atoms separately addressable 
by weak laser fields that drive Rabi oscillations.
The cavity mode, if excited, is also allowed to
escape  through conversion to photons at a partially transmitting
wall. This is the only decohering process.  The connection scheme 
for the relevant states is depicted in Fig.\ref{connections}.

%%%%%%%%%%%%%%%%%%%%%%%%%%%%%%%%%%%%%%%%%%%%%
%%%%%%%%%%%%%%%%%%%%%%%%%%%%%%%%%%%%%%%%%%%%%

The excitation of the cavity mode from the states in the chain that
are coupled thereto are the S links, denoted by fat double arrows. 
The thin double arrows are W links, and the double arrows leading to D
represent the cavity mode leakage to the outside. 
All W links come from the laser interactions, which are
three-fold: on atom \#2, an exactly tuned $0\! \leftrightarrow \!
2$ signal and an exactly tuned $1\! \leftrightarrow \! 2$ signal;
on atom \#1, just an exactly tuned $1\! \leftrightarrow \! 2$
signal, $180^{\circ}$ out of phase with the corresponding
signal on the other atom.

The structure shown in Fig.\ 4 has a general property in common
with the models discussed above; there is a bloc within which all
connections are weak,
\begin{equation}
 |1,1 \rangle _0\leftrightarrow | a \rangle _0  
\leftrightarrow | 1,0\rangle_0,
\label{decosubspace}
\end{equation}
characterized in Ref.\cite{beige1} as a DFS. 
The existence of this DFS is explained qualitively in the diagram of 
Fig.\ref{connections}.  In this model both the cavity mode coupling 
(thick arrows connecting states) and the cavity mode decay (labelled $D$) 
are of order ``strong''.   The cavity mode decay is necessary for obtaining 
the DFS, since in its absence the strongly coupled block by itself
\begin{equation}
|1,1 \rangle_2 \Leftrightarrow |s \rangle_1 \Leftrightarrow |2,2 \rangle_0
\end{equation}
has a zero eigenvalue. 
With the dissipative links, $D$, included through imaginary diagonal terms in the 
Hamiltonian matrix, all three eigenvalues are of ``strong'' level, and are thus
approximately decoupled from the three states 
$\{|1,1 \rangle _0, | a \rangle _0, | 1,0\rangle_0\}$ which form a  
DFS, as emphasized in Ref.\cite{beige1}.

Ref.\cite{beige1} proceeds from this observation to the construction
of a CNOT gate in which the transformation of states is effected
by means of accurately timed laser pulses. We now demonstrate
the action of an alternative mechanism for the controlled 
transformation of the states, following an exact analogue
of the protocol used in the first example in this paper.
In this mechanism the applied laser fields are slowly changed
in frequency, rather than turned on and off. 

To show this, we 
first write an effective Hamiltonian operating within the bloc
of Eq.\ (\ref{decosubspace}), as given in Ref.\cite{beige1}, 
but allowing the laser frequencies to be detuned by a small amount,
\begin{equation}
H_{\rm eff}\!=\!\frac{\Omega}{2}\left[ e^{i \Delta_A t} |1,0
\rangle _0 \langle a|_0 -  e^{i \Delta_B t} |a \rangle _0 \langle
1,1|_0 \right]+ {\rm h.c.}. \label{beigeham}
\end{equation}
Here $\Omega$ is a Rabi frequency, and 
the detunings of the $0\! \leftrightarrow\! 2$ and $
1\! \leftrightarrow\! 2$ lasers are given by $\Delta_{A}$ and
$\Delta_{B}$ respectively.
Transition probabilities in the indicated basis will be unaffected
by the transformation (acting in the atomic space only),
$\Psi'(t)= \exp(- i \Lambda t) \Psi (t)$, where
\begin{eqnarray}
2\Lambda \!&=&\! ({\Delta_A\!+\!\Delta_B})| 1,0 \rangle \langle
1,0 |\ +\ ({\Delta_B\!-\!\Delta_A})| a \rangle \langle a|
\nonumber
\\ && - \ ({\Delta_A\!+\! \Delta_B})| 1,1\rangle \langle 1,1 |,
\end{eqnarray}
giving the new effective Hamiltonian,
\begin{equation}
H'_{\rm eff}=\Lambda+\frac{\Omega}{2}[|1,0 \rangle _0 \langle
a|_0 -|a\rangle _0 \langle 1,1|_0]+ {\rm h.c.}
\label{detune}
\end{equation}

This is exactly the first example of the present paper, with the
five states reduced to three by taking only the symmetric states
in place of the (\#1,\#2) and (\#4,\#5) complexes. We may thus 
interchange the states, exactly as before, by adiabatically varying 
the detunings. For example, by adiabatically changing the combination 
${\Delta_A\!+\!\Delta_B}$, while keeping ${\Delta_A\!-\! \Delta_B}$  
constant, we can effect the interchange of the states $|1,1\rangle_0$ and 
$|1,0\rangle_0$, thus performing a CNOT operation. 

The full set of equations describing the evolution of all the states 
depicted in Fig.\ref{connections} is given below, where we have used 
the Hamiltonian from Ref.\cite{beige1} and added detuning as in 
Eq.\ref{detune}:

\begin{eqnarray}
\dot{c}_{|a\rangle_0} &=& 
-\frac{i \Omega}{2} \left( c_{|10\rangle_0} - c_{|11\rangle_0} 
+ c_{|22\rangle_0} \right) \nonumber\\
&& \quad\quad  -\frac{i}{2} \left( \Delta_A -\Delta_B \right)c_{|a\rangle_0},
\nonumber\\
\dot{c}_{|10\rangle_0} &=& 
-\frac{i \Omega}{2} \left( c_{|a\rangle_0} + \frac{1}{\sqrt{2}} c_{|2,0\rangle_0} 
+ c_{|s\rangle_0} \right) \nonumber\\
&& \quad\quad   - \frac{i}{2} \left( \Delta_A +\Delta_B \right)c_{|10\rangle_0},
\nonumber\\
\dot{c}_{|11\rangle_0} &=& 
-\frac{i \Omega}{2} c_{|a\rangle_0} 
+ \frac{i}{2} \left( \Delta_A +\Delta_B \right)c_{|11\rangle_0},
\nonumber\\
\dot{c}_{|s\rangle_0} &=&  
-\frac{i \Omega}{2}  c_{|10\rangle_0}+ \sqrt{2} g c_{|11\rangle_1},
\nonumber\\
\dot{c}_{|11\rangle_1} &=& 
\sqrt{2} g c_{|s\rangle_0} - \kappa c_{|11\rangle_1},
\nonumber\\
\dot{c}_{|s\rangle_1} &=& 
g \left( c_{|22\rangle_0} - \sqrt{2} c_{|11\rangle_2} \right) 
- \kappa c_{|s\rangle_1}, 
\nonumber\\
\dot{c}_{|22\rangle_0} &=& 
\frac{i \Omega}{2} \left( c_{|a\rangle_0} - \sqrt{2} c_{|20\rangle_0} \right) 
- g c_{|s\rangle_1},
\nonumber\\
\dot{c}_{|20\rangle_0} &=& 
- \frac{i \Omega}{2} \left( \frac{1}{\sqrt{2}} c_{|10\rangle_0} 
- \sqrt{2} c_{|22\rangle_0} \right) - g c_{|10\rangle_1}
\nonumber\\
\dot{c}_{|10\rangle_1} &=& 
g c_{|20\rangle_0} - \kappa c_{|10\rangle_1}, 
\nonumber\\
\dot{c}_{|11\rangle_2} &=& 
\sqrt{2} g c_{|s\rangle_1} - 2\kappa c_{|11\rangle_2}. 
\end{eqnarray}
In these equations, $g$ is the parameter setting the strong scale
depicted by the fat double-arrows in Fig.\ \ref{connections}, while
$\kappa$ sets the (strong) scale of the dissipative links $D$ to
the environment (decay of cavity modes).

%%%%%%%%%%%%%%%%%%%%%%%%%%%%%%%%%%%%%%%%%%%%%
\begin{figure}[t]
\begin{center}
\epsfig{file=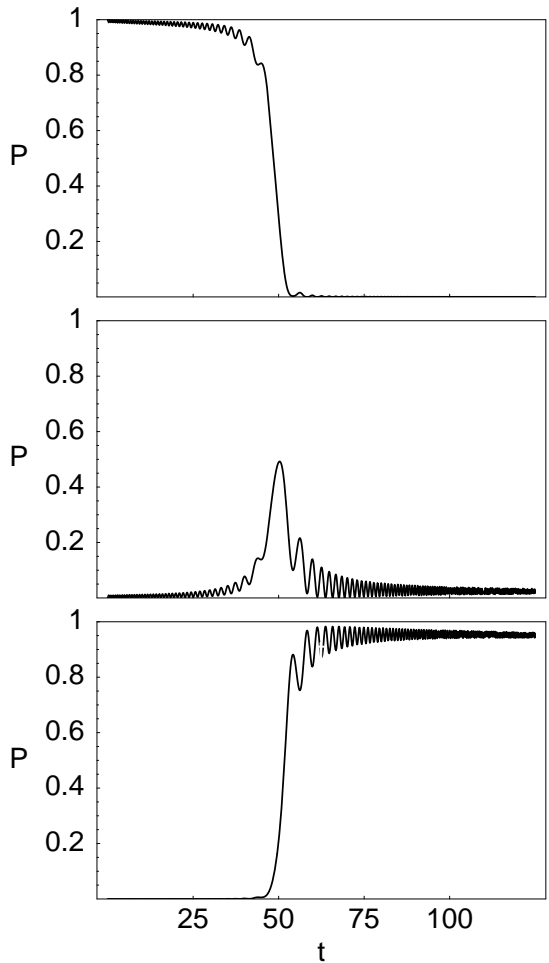,width=3.25in}
\caption{\label{detuning}
The top panel shows the survival probability of the initial state 
$| 10\rangle_0$, the center panel $|a\rangle_0$, and the lower panel 
$| 11\rangle_0$. The horizontal axis displays time in units of $\Omega^{-1}$.
We have taken the parameters $\kappa = g =2000 \Omega$,
$\Delta_A=(10-0.2t)\Omega$ and $\Delta_A=(10-0.2t)\Omega$.  The CNOT
opertation is implemented by adiabatically changing the detunings 
$\Delta_A + \Delta_B$ while keeping $\Delta_A - \Delta_B$ constant.}
\end{center}
\end{figure}  
%%%%%%%%%%%%%%%%%%%%%%%%%%%%%%%%%%%%%%%%%%%%%

We have solved this set of equation numerically, to demonstrate the 
interchange of states $|10\rangle_0 \leftrightarrow |11\rangle_0$.
We begin with a steady situation in which detuning parameters 
$\Delta_{A}$ and $\Delta_B$ are substantially greater than $\Omega$, 
and the initial state is (very stably) either $|1,1\rangle_0$ or 
$|1,0\rangle_0$. Then the detuning is manipulated, using simultaneous 
slow changes of both laser frequencies in order to interchange these two
states, in a process similar to those described in section \ref{spinchain}.  
The evolution is shown graphically in Fig.\ref{detuning}. 
By contrast, Ref.\cite{beige1} uses the perfectly tuned case 
$\Delta_{A,B} = 0$, and takes $\Omega = 0$ until a pulse turn-on time.
The pulse is then turned off at exactly the time for the
interchange $|1,0 \rangle _0\! \leftrightarrow\! |1,1\rangle _0 $
to have occurred under the influence of precession alone. In either
method, the transformation represents a CNOT gate, the states 
$|0,1\rangle_0$ and $|0,0\rangle_0$ being frozen due to the S-W effect. 
Note that our implementation does not require accurate timing of applied fields.

\section{Discussion and conclusion}
\label{conclusion}

We have shown that in systems with appropriate 
arrangements of strong (S) and weak (W)
couplings, variable potentials applied to a relatively small
number of sites can efficiently effect state permutations
for spin chains with pure-exchange coupling. Using the same
approach, we found a new way to implement a CNOT gate
in an atoms-in-a-cavity model discussed by previous 
authors \cite{beige1}.

A number of models similar to that of Ref.\
\cite{beige1} can be found in the recent literature
\cite{itano,block,beige2,facchi,pachos,jane}.  For many of
these cases, the generic S-W paradigm developed in the
present paper provides a unified basis for understanding
how isolated subspaces are generated.

As our final comment we note that
the S-W non-dissipative links, and decay-induced dissipation -- the 
two cases specifically studied here --
are not the only means by which
subspaces can become mutually isolated. For instance, Ref.\
\cite{stodolsky} and many subsequent works have discussed cases in
which rapid incoherent scattering can freeze a
system in a single state or subspace of states.

\section*{Acknowledgements}

NFB was supported Fermilab (operated by URA under DOE contract 
DE-AC02-76CH03000) and by NASA grant NAG5-10842,
RRV by the Australian Research Council and
YYYW by the DOE grant DE-FG02-84ER40163.


\begin{thebibliography}{99}
\bibitem{footnote1} Permutations are denoted by $\langle \cdots
\rangle$, where, e.g., $\langle123 \rangle \langle 4 \rangle
\langle 5 \rangle$ implies $1\to 2$, $2\to 3$, $3\to 1$.
\bibitem{footnote2}If the laser-induced
couplings and the cavity mode coupling were all 
of the same order of magnitude as the dissipation through
leakage D, then a system initially within the
subspace of Eq.\ (\ref{decosubspace})  would flow rapidly to the modes
at the edges of Fig.\ 4, and out the door D.
\end{thebibliography}
\end{document}